\documentclass[12pt]{article}
\usepackage{epsfig}

\renewcommand{\thefootnote}{\fnsymbol{footnote}}

\begin{document}

\title{
\begin{flushright}
\ \\*[-80pt] 
\begin{minipage}{0.2\linewidth}
\normalsize
arXiv:0708.3148 \\
YITP-07-47 \\
KUNS-2088 \\*[50pt]
\end{minipage}
\end{flushright}
{\Large \bf 
R-symmetry, supersymmetry breaking and \\
metastable vacua in global and local \\ 
supersymmetric theories
\\*[20pt]}}

\author{Hiroyuki~Abe$^{1,}$\footnote{
E-mail address: abe@yukawa.kyoto-u.ac.jp}, \ 
Tatsuo~Kobayashi$^{2,}$\footnote{
E-mail address: kobayash@gauge.scphys.kyoto-u.ac.jp} \ and \ 
Yuji~Omura$^{3,}$\footnote{E-mail address: omura@scphys.kyoto-u.ac.jp
}\\*[20pt]
$^1${\it \normalsize 
Yukawa Institute for Theoretical Physics, Kyoto University, 
Kyoto 606-8502, Japan} \\
$^2${\it \normalsize 
Department of Physics, Kyoto University, 
Kyoto 606-8502, Japan} \\
$^3${\it \normalsize 
Department of Physics, Kyoto University, 
Kyoto 606-8501, Japan} \\*[50pt]}

\date{
\centerline{\small \bf Abstract}
\begin{minipage}{0.9\linewidth}
\medskip 
\medskip 
\small
We study $N=1$ global and local supersymmetric theories with a 
continuous global $U(1)_R$ symmetry as models of dynamical 
supersymmetry (SUSY) breaking. We introduce explicit R-symmetry breaking 
terms into such models, in particular a generalized O'Raifeartaigh model.
Such explicit R-symmetry breaking terms can lead 
to a SUSY preserving minimum.
We classify explicit R-symmetry breaking terms 
by the structure of newly appeared SUSY stationary 
points as a consequence of the R-breaking effect, which could make 
the SUSY breaking vacuum metastable. 
We show that the R-breaking terms are basically divided into two 
categories. One of them does not generate a SUSY solution, or yields 
SUSY solutions that disappear in the case of supergravity when we 
tune a parameter so that the original SUSY breaking minimum becomes 
a Minkowski vacuum. 
We also show that the general argument by Nelson and Seiberg for a 
dynamical SUSY breaking still holds with a local SUSY except for a 
certain nontrivial case, and present concrete examples of the exception. 
\end{minipage}
}

\begin{titlepage}
\maketitle
\thispagestyle{empty}
\end{titlepage}

\tableofcontents
\thispagestyle{empty}
\clearpage

\renewcommand{\thefootnote}{\arabic{footnote}}
\setcounter{footnote}{0}
\setcounter{page}{1}

\section{Introduction}
Supersymmetric extensions of the standard model are promising 
candidate for the physics around TeV scale. Supersymmetry (SUSY) can 
stabilize the huge hierarchy between the weak scale and the 
Planck scale, and supersymmetric models with R-parity have the 
lightest superparticle as a good candidate for the dark matter. 
In addition, the minimal SUSY standard model realizes 
the unification of three gauge couplings at a scale 
$M_{GUT} \sim 2 \times 10^{16}$ GeV, which may suggest some 
underlying unified structure in the nature.

In our real world, the SUSY must be broken with 
certain amount of the gaugino and scalar masses. The dynamical 
SUSY breaking has a big predictability of the structure 
of such SUSY particles. It was shown by Nelson and 
Seiberg (NS)~\cite{Nelson:1993nf} that a global $U(1)_R$ symmetry is 
necessary for a spontaneous F-term SUSY breaking at 
the ground state of generic models with a global SUSY. 
This predicts an appearance of massless Goldstone mode, R-axion, 
in dynamically SUSY breaking models with nonvanishing 
Majorana gaugino masses which breaks $U(1)_R$ symmetry.\footnote{
See for recent works on R-symmetry breaking, e.g. 
Refs.~\cite{Shih:2007av,Intriligator:2007py,Ferretti:2007ec,Essig:2007xk,
Cho:2007yn,Abel:2007jx,Ray:2007wq} 
and references therein.}

Recently, it has been argued by Intriligator, Seiberg and Shih 
(ISS)~\cite{Intriligator:2007py} that the SUSY breaking 
vacuum we are living can be metastable for avoiding the 
light R-axion and also obtaining gaugino masses, 
and that such situation can be realized by a tiny size of 
explicit $U(1)_R$ breaking effects, whose representative 
magnitude is denoted by $\epsilon$.
Such explicit R-symmetry breaking terms can lead to 
a SUSY minimum, but such newly appeared 
SUSY minimum could be far away from the
SUSY breaking minimum, which is found in 
the R-symmetric model without explicit R-symmetry 
breaking terms.
Furthermore, such R-symmetry breaking terms 
would not have significant effects on the 
original SUSY breaking minimum, because 
R-symmetry breaking terms are tiny.
The distance between the original SUSY breaking 
minimum and the newly appeared SUSY preserving 
minima may be estimated as ${\cal O}(1/\epsilon)$ 
in the field space.
Thus, if R-symmetry breaking terms, i.e., 
the size of $\epsilon$, are sufficiently small, 
the original SUSY breaking minimum would be a 
long-lived metastable vacuum.

On the other hand, an introduction of gravity into 
SUSY theories requires that the SUSY 
must be a local symmetry, i.e., supergravity. In supergravity, 
the structure of the scalar potential receives a gravitational 
correction, and also the background geometry of our spacetime 
is determined by the equation of motion depending upon the 
vacuum energy. 
In the above global SUSY model with metastable SUSY breaking 
vacuum, some fields have large vacuum values at the 
SUSY preserving vacuum.
In such a case, supergravity effects might be sizable.
Another important motivation to consider supergravity is
to realize the almost vanishing vacuum energy.
The global SUSY model always has positive vacuum energy 
at the SUSY breaking minimum.
Supergravity effects could realize almost vanishing 
vacuum energy.

F-flat conditions have supergravity corrections. 
Thus, the supergravity model with global $U(1)_R$ symmetry 
would have different aspects from the global SUSY model.
Furthermore, adding R-symmetry breaking terms 
would have different effects between global and 
local SUSY theories.
Here we study in detail generic aspects of 
global and local SUSY theories with R-symmetry 
and generic behaviors caused by adding explicit 
R-symmetry breaking terms.
That is,  we reconsider the above argument for the 
dynamical SUSY breaking and its metastability by 
NS and ISS comparing global and local SUSY theories.

The important keypoint is to realize 
the almost vanishing vacuum energy.
That is impossible in the SUSY breaking vacuum of 
global SUSY models, and that is a challenging 
issue in supergravity models.
The vacuum energy may be tuned to vanish, e.g., by 
the constant superpotential term, which 
is a sizable R-symmetry breaking term.
That would affect all of vacuum structure such as 
metastability of SUSY breaking vacua and 
presence of SUSY preserving vacua.
Here we study this vacuum structure by 
using several concrete models, where we start R-symmetric 
models and add certain classes of R-symmetry breaking 
terms such that the vanishing vacuum energy is 
realized.

The following sections are organized as follows. 
In Sec.~\ref{sec:global-SUSY}, we study a structure of dynamical 
SUSY breaking in R-symmetric models with global SUSY. 
We consider the generalized O'Raifeartaigh (OR) 
model~\cite{O'Raifeartaigh:1975pr} following \cite{Intriligator:2007py}.
We introduce 
explicit R-breaking terms into the model and analyze in detail 
the newly appeared SUSY vacua as a consequence of the R-symmetry 
breaking effects. We also examine the stability of the original 
SUSY breaking vacuum under such R-breaking terms. 

In Sec.~\ref{sec:rsugra}, we consider supergravity models with 
R-symmetry. We extend the argument by NS to the local SUSY theories 
and study the supergravity OR model. 
In this section, we also show a special SUSY 
stationary point, which does not obey the NS condition, and the 
associated SUSY breaking vacuum in a certain class of R-symmetric 
supergravity models. We introduce explicit R-breaking terms into 
the supergravity OR model in Sec.~\ref{sec:r-breaking-sugra} 
and classify them by the consequent SUSY solutions. 

In Sec.~\ref{sec:typea}, we study the case with R-symmetry breaking 
terms (A-type) which might not cause a metastability of SUSY breaking 
minimum, because corresponding SUSY vacua disappear when we set the 
vacuum energy at the SUSY breaking minimum vanishing. 
On the other hand, in Sec.~\ref{sec:typeb}, we show that another 
class of R-symmetry breaking terms (B-type) makes SUSY breaking 
minimum metastable. 
Sec.~\ref{sec:conclusion} is devoted to conclusions. 
In Appendix~\ref{app:rmass}, we show some general features 
of R-axion masses, and find that the special SUSY 
solution exhibited in Sec.~\ref{sec:rsugra} is at best a 
saddle point solution.

\section{Global supersymmetric theory}
\label{sec:global-SUSY}

\subsection{R-symmetric model}
First, we review briefly the argument by Nelson and 
Seiberg~\cite{Nelson:1993nf} in R-symmetric global SUSY models. 
Let us consider the global SUSY model with 
$n$ chiral superfields $z_I$ ($I=1,\ldots,n$) and 
their superpotential $W(z_I)$.
In the case of global SUSY, F-flat conditions are 
determined by 
\begin{eqnarray}
W_{z_I} &=& 0, 
\label{eq:fflat}
\end{eqnarray}
where $W_{z_I}=\partial_{z_I}W$.
Hereafter we use a similar notation for derivatives 
of functions $H(X)$ by fields $X$ as $H_X$.
The conditions (\ref{eq:fflat}) are $n$ complex equations for 
$n$ complex variables, and these can have a solution for generic 
superpotential. 

Now, we consider global SUSY models with a continuous global 
$U(1)_R$ symmetry and a nonvanishing superpotential.
Since the superpotential has the R-charge $2$, 
there exists at least one field with a nonvanishing R-charge. 
Suppose that the $n$-th component $z_n$ is such a field with 
the nonvanishing R-charge, $q_{z_n} \ne 0$. 
Then, in the following field basis 
\begin{eqnarray}
\chi_i &=& \frac{z_i}{z_n^{q_{z_i}/q_{z_n}}}, 
\qquad (q_{\chi_i}=0), 
\nonumber \\
Y &=& z_n, 
\qquad (q_Y=q_{z_n}\ne 0), 
\label{eq:rbasis1}
\end{eqnarray}
where $i=1,2,\ldots,n-1$, 
the superpotential can be written as 
\begin{eqnarray}
W_{NS} &=& Y^{2/q_Y}\zeta(\chi_i). 
\label{eq:rsp1}
\end{eqnarray}
Then the F-flat conditions (\ref{eq:fflat}) are 
split into two pieces, 
\begin{eqnarray}
(2/q_Y)Y^{2/q_Y-1}\zeta(\chi_i) &=& 0, 
\label{eq:fflatx} \\
Y^{2/q_Y} \partial_{\chi_j} \zeta(\chi_i) &=& 0. 
\label{eq:fflatphi}
\end{eqnarray}
When we look for an R-symmetry breaking vacuum, 
$\langle Y \rangle \ne 0$, 
these conditions are equivalent to 
\begin{eqnarray}
\zeta(\chi_i) &=& 0, \qquad 
\partial_{\chi_j} \zeta(\chi_i) \ = \ 0, 
\label{eq:globalsusyvc}
\end{eqnarray}
which are $n$ complex equations for $n-1$ complex variables, 
that is, these are {\it over-constrained} conditions. 
These cannot be satisfied at the same time for a generic 
function $\zeta(\chi_i)$, and the SUSY can be broken. 
This is an observation by Nelson and Seiberg~\cite{Nelson:1993nf} 
that the existence of an R-symmetry is the necessary condition 
for a dynamical SUSY breaking, and is also the sufficient condition 
if the R-symmetry is spontaneously broken, $\langle Y \rangle \ne 0$.

However, the scalar potential, which 
is obtained from the superpotential (\ref{eq:rsp1}) and  
the K\"ahler potential $K(|Y|,\chi_i,\bar\chi_i) $, 
is found to have the global minimum at $Y=0$, unless 
the K\"ahler potential $K(|Y|,\chi_i,\bar\chi_i) $ is 
non-trivial.
Thus, SUSY is not broken dynamically with 
the NS superpotential (\ref{eq:rsp1}).

The O'Raifeartaigh model~\cite{O'Raifeartaigh:1975pr} 
is a good example of R-symmetric SUSY models, where SUSY 
is spontaneously broken.
Its generalization is shown in Ref.~\cite{Intriligator:2007py} 
as the generalized OR model, which 
has the following superpotential,
 \begin{eqnarray}
W_{OR} &=& \sum_a g_a(\phi_i)\,X_a, 
\label{eq:orgsp}
\end{eqnarray}
where $a=1,2,\ldots,r$ and $i=1,2,\ldots,s$, and 
the numbers of fields are constrained as $r>s$. 
Their R-charges are assigned as $q_{X_a}=2$ and $q_{\phi_i}=0$, 
and $g_a(\phi_i)$ is a function of $\phi_i$. 
In this model, $F$-flat conditions for $X_a$ are just given by 
\begin{eqnarray}
\partial_{X_a} W &=& g_a(\phi_i) \ = \ 0.
\label{eq:orgfflat}
\end{eqnarray}
These are $r$ complex equations for $s$ complex valuables, 
that is, these are {\it over-constrained} conditions 
for $r>s$. 
Therefore, there is no SUSY solution satisfying 
(\ref{eq:orgfflat}) for generic functions $g_a(\phi_i)$ 
with $r>s$. 
The superpotential of the generalized OR model (\ref{eq:orgsp})
is a specific form of the NS superpotential 
(\ref{eq:rsp1}).
In the generalized OR model, SUSY is always spontaneously broken 
independently of whether R-symmetry is spontaneously broken 
or not, or the fields $X_a$ develop nonvanishing vacuum expectation 
values or not.

The simplest OR model is the model with $r=1$ and $s=0$, 
and has the superpotential 
\begin{eqnarray}
 & & W_{(OR)_1}=fX_1 ,
\nonumber
\end{eqnarray}
where $f$ is a constant.
Obviously, SUSY is spontaneously broken in this model, 
because $W_{X_1}=f$.
The basic O'Raifeartaigh model corresponds 
to the model with $r=2$ and $s=1$, and 
$g_1(\phi)=f + \frac12 h\phi^2$ and $g_2(\phi)=m\phi$, 
and has the following superpotential, 
\begin{eqnarray}
W_{(OR)_{basic}} &=& (f+\frac{1}{2} h \phi^2) X_1 +m \phi X_2. 
\label{eq:oorsp}
\end{eqnarray} 
 The model has only a SUSY breaking pseudo-moduli space, 
\begin{eqnarray}
\phi &=& X_2 \ = \ 0, \qquad 
X_1 \ : \ \textrm{undetermined}, 
\label{eq:or1pms}
\end{eqnarray}
with $W_{X_1}=f$ as a global minimum of the potential. 
When integrating out heavy modes $X_2$ and $\phi$, we 
obtain $W_{(OR)_1}$ as an effective superpotential.
However, the flat direction along $X_1$ is lifted at the one-loop level 
by integrating out $\phi$, and the SUSY breaking 
vacuum in the quantum corrected OR model is given by
\begin{eqnarray}
\phi &=& X_2 \ = \ X_1 \ = \ 0. 
\label{eq:qor1vac}
\end{eqnarray}

These simple models suggest that the tadpole term of $X_a$ 
is important for SUSY breaking.
Indeed, we can show by simple discussion that 
non-vanishing terms of  $g_a(\phi_i)$ at $\phi_i=0$
are sources of SUSY breaking.
We assume that $g_a(\phi_i)$ are non-singular functions.
Then, we can always rewrite the superpotential (\ref{eq:orgsp}) as 
\begin{eqnarray}
W_{OR} &=& \sum_a f_a X_a +\sum_a \check{g}_a(\phi_i) X_a 
\nonumber \\ &=& 
\tilde{f} \tilde{X}_1 
+\sum_a \tilde{g}_a(\phi_i) \tilde{X}_a, \qquad 
(\tilde{g}_a(0) \ = \ 0), 
\label{eq:orgsprd}
\end{eqnarray}
where 
$f_a=g_a(0)$, $\check{g}_a(\phi_i)=g_a(\phi_i)-f_a$, 
$\tilde{X}_a=U_{ab}X_b$, 
$\tilde{g}_a(\phi_i)=\check{g}_a U^\dagger_{ab}$ 
and $U_{ab}$ is a constant unitary matrix defined by 
$f_a U^\dagger_{ab}=\tilde{f}_b=(\tilde{f},0,\ldots,0)$. 
In the following, we will frequently use this basis of 
fields and omit the tildes to simplify the notation. 
In this basis, the F-flat conditions for $X_a$, 
Eq.~(\ref{eq:orgfflat}), are written by 
\begin{eqnarray}
W_{X_a} &=& 
\ g_a(\phi_i)-\delta_{a1}f  
\ = \ 0 .
\label{eq:orgfflatsugrard}
\end{eqnarray}
Together with 
$W_{\phi_i}=\sum_a X_a \partial_{\phi_i}g_a(\phi_i)=0$, 
we find that, if $f=0$, there is a solution $X_a=\phi_i=0$ 
and SUSY is not broken. 
Then it is obvious in the field basis (\ref{eq:orgsprd}) 
that a nonvanishing $f$ is the source of dynamical 
SUSY breaking in the generalized OR model.

In the generalized OR model with the above field basis, 
the field $X_1$ plays a special role, 
while each of $X_a$ ($a \neq 1$) has the qualitatively same 
character as others $X_b$ ($b \neq 1$).
Thus, the simple model with $r=2$ and $s=1$, 
and the superpotential,
\begin{eqnarray}
W_{(OR)_2}=(f+g_1(\phi))X_1+g_2(\phi) X_2,
\nonumber
\end{eqnarray}
shows qualitatively generic aspects of 
the generalized OR model.
Its scalar potential is written as 
\begin{eqnarray}
V&=& |f+g_1(\phi)|^2+|g_2(\phi)|^2+|W_{\phi}|^2,
\nonumber 
\end{eqnarray}
and stationary conditions are obtained as 
\begin{eqnarray}
V_{X_1}&=&\overline {W_{\phi}}g'_1(\phi) =0, \nonumber \\
V_{X_2}&=&\overline {W_{\phi}}g'_2(\phi) =0, \nonumber \\
V_{\phi}&=&\overline {W_{\phi}}W_{\phi \phi}
+(\overline f+ \overline {g_1(\phi)})g'_1(\phi) 
+ \overline {g_2(\phi)}g'_1(\phi) =0, 
\nonumber 
\end{eqnarray}
where $g'_a(\phi)=dg_a(\phi)/d\phi$ and $W_\phi = \sum_a X_a g'_a(\phi)$.
Unless $W_{\phi}$ does not vanish, we would have 
{\it over-constrained} conditions, i.e., 
$g'_1(\phi) = g'_2(\phi) =0$ for generic functions.
Thus, in general, the solution of the above 
stationary conditions corresponds to 
\begin{eqnarray}
 & & W_{\phi} = X_1 g'_1(\phi) + X_2g'_2(\phi) =0 ,
\nonumber \\
 & & (\overline f+ \overline {g_1(\phi)})g'_1(\phi) 
+ \overline {g_2(\phi)}g'_1(\phi) =0. 
\label{eq:or2-min}
\end{eqnarray}
The latter is the condition to fix $\phi$.
For a fixed value of $\phi$, a ratio between 
$X_1$ and $X_2$ is fixed by the former condition, 
but the linear combination 
\begin{eqnarray}
X_1g'_2(\phi)-X_2g'_1(\phi), 
\label{eq:flatdirgen}
\end{eqnarray}
remains undetermined.
That is the pseudo-flat direction, and 
would be lifted by loop effects.
Similarly we can discuss models with 
several fields $X_a$ and $\phi_i$ ($r>s$).

\subsection{Explicit R-symmetry breaking and 
metastable vacua}
In order to have Majorana gaugino masses in addition to soft scalar 
masses, the R-symmetry must be broken spontaneously or explicitly 
at the SUSY breaking minimum we are living. 
On the other hand, as shown in the previous section, 
the NS argument requires an exact R-symmetry 
for the dynamical SUSY breaking. Then, an appearance of 
an unwanted massless Goldstone mode, an R-axion, is inevitable 
in such R-symmetry breaking minimum. Does this mean the dynamical 
SUSY breaking is phenomenologically disfavored ? 

Recently, it has been argued by Intriligator, Seiberg and 
Shih~\cite{Intriligator:2007py} that our world must reside in a 
metastable state, in order to avoid the above conflict between 
gaugino masses and the R-axion. The arguments are as follows. 
Consider a theory with an approximate R-symmetry which has a small 
R-symmetry breaking parameter $\epsilon$. 
In the limit $\epsilon \to 0$, the R-symmetry 
becomes exact, and the theory possesses a SUSY breaking 
ground state due to the NS argument. For a nonzero 
but tiny parameter $\epsilon$, this SUSY breaking minimum 
still remains as a local minimum of the potential, although there 
appear SUSY ground states somewhere in the field space 
due to explicit R-symmetry breaking effects. As long as 
the parameter $\epsilon$ is small enough, the separation between 
the SUSY breaking minimum and the supersymmetric vacua 
is large, and the former can be a long-lived metastable vacuum. 
These facts were exhibited by ISS based on the O'Raifeartaigh model 
as a simple example of dynamical SUSY breaking model 
with R-symmetry. 
Indeed, such O'Raifeartaigh-type model can be realized 
in some region of the moduli space of SUSY Yang-Mills 
theories~\cite{Intriligator:2006dd}. 

Here following the discussion by ISS we study generic 
aspects of explicit R-symmetry breaking terms, 
and SUSY preserving vacua.
We also classify  explicit R-symmetry breaking terms 
in global SUSY models.
In addition, we discuss metastability.

The simplest R-symmetry breaking term is the 
constant term $W_{R\!\!\!\!/} = c$, but 
the constant term does not play any role in global SUSY theory.
Thus, we do not discuss about adding constant term in this section.
It is obvious that when we add any R-symmetry breaking term 
$W_{R\!\!\!\!/}(Y,\chi)$ to the NS superpotential 
(\ref{eq:rsp1}), that can relax {\it over-constrained} 
conditions and F-flat conditions can have SUSY 
solutions.

The generalized OR model has richer structure 
in explicit R-symmetry breaking terms.
To see such structure, we consider the 
generalized OR model with 
three types of typical R-symmetry breaking terms, 
i) a function including only 
$\phi_i$ fields $W_{R\!\!\!\!/}=w(\phi)$, 
ii) a function including only $X_a$ ($a \neq 1$), 
$W_{R\!\!\!\!/}=w(X_a)$,
and iii) a function including only $X_1$, 
$W_{R\!\!\!\!/}=w(X_1)$.
The first type of R-symmetry breaking terms 
$W_{R\!\!\!\!/}=w(\phi)$ do not change 
F-flat conditions for $X_a$, i.e., 
$\partial_{X_a} W = f\delta_{a1}+g_a(\phi_i)=0$.
Hence, there is no SUSY solution.
 
For the second type of R-symmetry breaking terms 
$W_{R\!\!\!\!/}=w(X_a)$ ($a\neq 1$), F-flat conditions 
are obtained as 
\begin{eqnarray}
W_{X_1} &=& f+g_1(\phi_i) \ = \ 0, 
\nonumber \\
W_{X_a} &=& g_a(\phi_i)+w_{X_a}(W_a) \ = \ 0 
\qquad {\rm for ~~} a\neq 1, \nonumber \\
W_{\phi_i} &=& \sum_a X_a 
\partial_{\phi_i} g_a(\phi_i) \ = \ 0.
\nonumber
\end{eqnarray}
Thus, if $w_{X_a}(W_a) \neq 0$ for all of 
$X_a$, {\it over-constrained} conditions 
can be relaxed and a SUSY solution 
can be found.
If all of $\phi_i$ vanish, we have $g_1(\phi_i)=0$ 
and the condition $W_{X_1}=0$ can not be 
satisfied.
Hence, the SUSY minimum, which 
appears by adding $W_{R\!\!\!\!/}=w(X_a)$ ($a\neq 1$), 
corresponds to the point, where some of $\phi_i$ 
develop nonvanishing vacuum expectation values.

For the third type of R-symmetry breaking terms 
$W_{R\!\!\!\!/}=w(X_1)$, F-flat conditions are 
obtained as 
\begin{eqnarray}
W_{X_1} &=& f+g_1(\phi_i) +\partial_{X_1} w(X_1) \ = \ 0, 
\nonumber \\
W_{X_a} &=& g_a(\phi_i) \ = \ 0 
\qquad {\rm for ~~} a\neq 1, \nonumber \\
W_{\phi_i} &=& \sum_a X_a 
\partial_{\phi_i} g_a(\phi_i) \ = \ 0.
\nonumber
\end{eqnarray}
If $r=s+1$, the {\it over-constrained} conditions 
can be relaxed.
In this case, the point $\phi_i=0$ for all of $i$ 
can be a solution for $W_{X_a}=0 $ for $a \neq 1$.
Furthermore, the conditions,
\begin{eqnarray}
 & & f+\partial_{X_1} w(X_1) \ = \ 0, \qquad 
\sum_a X_a \partial_{\phi_i} g_a(\phi_i) \ = \ 0,
\nonumber
\end{eqnarray}
should be satisfied.

When R-symmetry breaking terms 
include $X_1$ and $X_a$ ($a \neq 1$), 
{\it over-constrained} conditions 
can be relaxed and a solution 
for F-flat conditions would correspond 
to $\phi_i \neq 0$ for some of $\phi_i$.

The SUSY breaking minimum is found 
in the generalized OR model without explicit 
R-symmetry breaking terms, as discussed 
in the previous subsection.
As discussed above, 
SUSY vacua can appear,
when we add the definite form of 
explicit R-symmetry breaking terms 
to the generalized OR model.
Thus, the previous SUSY breaking minimum 
is a metastable vacuum, 
if such R-symmetry breaking effects are 
small around the SUSY breaking minimum 
and the SUSY breaking vacuum itself is 
not destabilized by such 
R-symmetry breaking terms.
 
As an illustrating example, we consider 
the basic OR model (\ref{eq:oorsp}) 
with explicit R-symmetry breaking terms.
ISS introduced an explicit R-symmetry breaking term in the 
superpotential\footnote{See also Ref.~\cite{Intriligator:2007cp}.}, 
$W=W_{(OR)_{basic}}+W_{R\!\!\!\!/}$, where 
\begin{eqnarray}
W_{R\!\!\!\!/} &=& \frac{1}{2}\epsilon m X_2^2. 
\label{eq:rbrkex1}
\end{eqnarray}
In this case, there appears a SUSY minimum, 
\begin{eqnarray}
\phi &=& \sqrt{-\frac{2f}{h}}, \qquad 
X_2 \ = \ -\frac{1}{\epsilon} \phi, \qquad 
X_1 \ = \ \frac{m}{\epsilon h}, 
\nonumber
\end{eqnarray}
which is far away from the (local) SUSY breaking 
minimum (\ref{eq:qor1vac}) 
for a sufficiently small $\epsilon \ll 1$. 
In addition, the SUSY breaking minimum is 
not destabilized by the above R-symmetry breaking 
term (\ref{eq:rbrkex1}).
Then the original 
SUSY breaking vacuum (\ref{eq:qor1vac}) becomes metastable 
which can be parametrically long-lived for $\epsilon \ll 1$.

Instead, if we consider the following R-breaking term~\cite{Dine:2006gm} 
\begin{eqnarray}
W_{R\!\!\!\!/} &=& \frac{1}{2}\epsilon m X_1^2, 
\label{eq:rbrkex2}
\end{eqnarray}
the newly appeared SUSY point is found as 
\begin{eqnarray}
\phi &=& X_2 \ = \ 0, \qquad 
X_1 \ = \ -\frac{f}{\epsilon m}. 
\nonumber
\end{eqnarray}
In this case, the pseudo-moduli space (\ref{eq:or1pms}) 
disappears at the tree level. 
However, the SUSY 
breaking point (\ref{eq:qor1vac}) remains as a local 
minimum due to the one-loop mass for $X_1$, but becomes 
metastable. Then the situation is similar to the above example. 
We easily find that any R-breaking terms 
which consist of only $\phi$ do not restore SUSY.

Now, let us study whether the 
SUSY breaking minimum, which is found without 
R-symmetry breaking terms, is destabilized by adding 
R-symmetry breaking terms.
We consider the generalized OR model with $(r=2,s=1)$, i.e., 
$W_{(OR)_2}$,  
whose stationary conditions (\ref{eq:or2-min}) are studied in the 
previous subsection.
Their solutions are denoted by $X_a=X^{(0)}_a$ and 
$\phi=\phi^{(0)}$.
First, we add a small R-symmetry breaking term, 
$W_{R\!\!\!\!/} = \epsilon w(X_2)$, which 
depends only on $X_2$.
Then, the scalar potential is written as 
\begin{eqnarray}
 & & V=|f+g'_1(\phi)|^2+|g_2(\phi)+\epsilon w'(X_2)|^2 +
|W_\phi|^2,
\nonumber
\end{eqnarray}
where $W_\phi=X_1g'_1(\phi)+X_2g'_2(\phi)$.
In addition, we assume that the stationary conditions of $V$ 
are satisfied by $X_a=X^{(0)}_a+\delta X_a$ and 
$\phi=\phi^{(0)} + \delta \phi$, and that all of $\delta X_a$ and 
$\delta \phi$ are of ${\cal O} (\epsilon)$.
For example, the stationary condition along $\phi$, 
$V_\phi =0$, gives the following condition,
\begin{eqnarray}
 & & \left( \sum_a |g'_a(\phi^{(0)})|^2 +
\sum_a(\overline f\delta_{a1}+\overline{g_a(\phi^{(0)})})g''_a(\phi^{(0)}))
\right) \delta \phi 
+  \epsilon g'_2(\phi^{(0)}) \ \overline {w'(X_2^{0})} =0,
\nonumber
\end{eqnarray}
where we have used the stationary conditions (\ref{eq:or2-min})  
at $X_a=X_a^{(0)}$ and $\phi = \phi^{(0)}$.
This is the equation to determine $\delta \phi$.
The stationary condition along $X_1$, $V_{X_1}=0$, reduces to 
\begin{eqnarray}
 & & g'_1(\phi^{(0)}) \ \overline{\delta W_\phi} =0,
\nonumber
\end{eqnarray}
where 
\begin{eqnarray}
 & & \delta W_\phi = \sum_a g'_a(\phi^{(0)})\delta X_a 
+ \sum_aX_a^{(0)}g''_a(\phi)\delta \phi .
\nonumber
\end{eqnarray}
Thus, this shows a relation among $\delta X_a$ and $\delta \phi$ 
unless $g'_1(\phi^{(0)})=0$.
On the other hand, the stationary condition along 
$X_2$, $V_{X_2}=0$, leads to the following equation,
\begin{eqnarray}
 & & \epsilon w''(X_2^{(0)}) \ \overline{g_2(\phi^{(0)})}=0.
\nonumber
\end{eqnarray}
This is not an equation among $\delta X_a$ and $\delta \phi$, 
but implies that the stationary condition is destabilized 
unless $w''(X_2^{(0)}) \ \overline{g_1(\phi^{(0)})}  =0$.
In the above basic O'Raifeartaigh model, we have 
$g_1(\phi^{(0)})=0$.
Thus, the SUSY breaking minimum is not destabilized by 
adding the mass term of $X_2$, $w(X_2) =\frac12 mX_2$, i.e., 
$w''(X_2) \neq 0$ at $X_2=0$.

Now, let us add an R-symmetry breaking term, 
$W_{R\!\!\!\!/} = \epsilon w(X_1)$, which 
depends only on $X_1$.
Similarly, we can examine stationary conditions of 
the scalar potential, 
\begin{eqnarray}
 & & V=|f+g'_1(\phi)+\epsilon w'(X_1)|^2+|g_2(\phi)|^2 +
|W_\phi|^2.
\nonumber
\end{eqnarray}
The stationary conditions along $X_2$ and $\phi$ 
give an equation to determine $\delta \phi$ and 
a relation among $\delta X_a$ and $\delta \phi$.
However, the stationary condition along $X_1$, 
$V_{X_1}=0$, leads to 
\begin{eqnarray}
 & & w''(X_1^{(0)})  
\left( \overline{f}+\overline{g_1(\phi^{(0)})}\right)  =0.
\nonumber
\end{eqnarray}
If this condition is not satisfied, 
the stationary condition at the 
SUSY breaking vacuum is destabilized.
Indeed, the basic O'Raifeartaigh model has 
$f+g_1(\phi)=f$ at $\phi =0$.
Thus, when we add the mass term of $X_1$, 
$w(X_1)=\frac12 mX_1^2$, i.e., $w'' \neq 0$, 
the SUSY breaking minimum become destabilized 
at the tree level as shown above.
Note that this kind of destabilization would be 
related to the existence of the flat direction 
(\ref{eq:flatdirgen}) in the OR model with global SUSY.\footnote{
Such flat direction would be lifted by supergravity effects.}

The above discussion shows that 
adding generic R-symmetry breaking terms 
can destabilize the SUSY breaking minimum, which 
is found in the model without such explicit 
R-symmetry breaking terms.
In order to realize metastability of 
the original SUSY breaking minimum, 
we need a certain type of R-symmetry breaking terms. 
Alternatively, loop-effects would be helpful not to 
destabilize the original SUSY breaking minimum 
by R-symmetry breaking terms.

\section{R-symmetry in supergravity}
\label{sec:rsugra}

In the previous section, based on the argument by ISS, 
we have shown that some sort of explicit R-symmetry breaking 
terms can restore SUSY, and the original SUSY 
breaking vacuum can become metastable when 
a certain (but not generic) class of explicit R-symmetry breaking terms 
are added and/or loop effects stabilize 
the original SUSY breaking minimum. 
The metastable minimum 
can be parametrically long-lived if the coefficient of the 
R-breaking term is sufficiently small with which the SUSY 
ground state is far from the metastable state in the field space. 

This argument has been performed in a decoupling limit of gravity. 
As we find in the above discussion, however, we have to treat a large 
distance between some separated minima in the field space. This may 
imply that large vacuum values of some fields might be involved in 
the analysis, where supergravity effects could become sizable. 
Moreover, in global SUSY, the SUSY breaking minima 
always have a positive vacuum energy with the magnitude of the 
SUSY breaking scale, which never satisfies the observation 
that the vacuum energy almost vanishes. 
In such a sense, we would be forced to consider supergravity.

Note that, even in supergravity, it is often a hard task to tune 
the vacuum energy at the stationary points of the scalar potential 
to be almost vanishing. This might require a large R-symmetry breaking 
effect specialized to supergravity, i.e., a constant term in the 
superpotential~\cite{Bagger:1994hh}. The existence of such a special 
R-symmetry breaking term could also affect the ISS argument of 
metastability. 
Loop effects have contributions to the vacuum energy.
Here we assume that such loop effects are subdominant, 
and we tune our parameters such that we realize $V \approx 0$ 
at the tree level.
Hereafter we use the unit with $M_{Pl}=1$, where 
$M_{Pl}$ denotes the reduced Planck scale.

\subsection{Nelson-Seiberg argument}

In this subsection, we study the NS argument 
within the framework of supergravity theory.
In the case of supergravity, F-flat conditions (\ref{eq:fflat}) are 
modified as 
\begin{eqnarray}
D_IW &\equiv& W_I+K_I W \ = \ 0, 
\nonumber
\end{eqnarray}
where $K$ denotes the K\"ahler potential, $K(|Y|,\chi_i,\bar \chi_i)$. 
In the field basis (\ref{eq:rbasis1}) with the superpotential 
(\ref{eq:rsp1}), these are written as 
\begin{eqnarray}
D_{\chi_i}W &=& Y^{2/q_Y}(\zeta_i+K_i \zeta) \ = \ 0, 
\nonumber \\
D_Y W &=& (2/q_Y+YK_Y)X^{2/q_Y-1}\zeta \ = \ 0.
\nonumber
\end{eqnarray}
Then, we find the following two candidates of 
R-breaking SUSY solutions 
in supergravity, 
\begin{eqnarray}
\zeta_i &=& 0, \qquad 
\zeta \ = \ 0, 
\label{eq:susyvc1}
\end{eqnarray}
and 
\begin{eqnarray}
D_{\chi_i} \zeta &=& \zeta_i+K_i \zeta \ = \ 0, \qquad 
2/q_Y+YK_Y \ = \ 0. 
\label{eq:susyvc2}
\end{eqnarray}

The first conditions (\ref{eq:susyvc1}) contain $n$ complex equations 
for $n-1$ complex variables, and the situation is the same as the case 
of global SUSY (\ref{eq:globalsusyvc}), that is, the solution 
does not exist for a generic function $\zeta$. 
On the other hand, the second conditions (\ref{eq:susyvc2}) are $n$ 
complex equations for $n$ complex variables which can have a solution. 
This corresponds to a SUSY stationary point specialized to 
R-symmetric supergravity.

In this subsection, we analyze the special SUSY stationary 
solution (\ref{eq:susyvc2}) which appears due to purely the 
supergravity effect and does not obey the NS condition. 
Then, in the following we assume that there is a solution for 
\begin{eqnarray}
2/q_Y+YK_Y &=& 0. 
\label{eq:susyvc2:2}
\end{eqnarray}

For instance, if the K\"ahler potential is given by 
\begin{eqnarray}
K &=& \sum_{n_Y=1}c_{n_Y}|Y|^{2n_Y}
+\hat{K}(\chi_i,\bar\chi_i), 
\label{eq:kxpol}
\end{eqnarray}
the condition (\ref{eq:susyvc2:2}) becomes 
\begin{eqnarray}
2/q_Y+\sum_{n_Y=1}n_Y c_{n_Y}|Y|^{2n_Y} &=& 0. 
\nonumber
\end{eqnarray}
Then, we need at least one negative value of $\{ c_{n_Y},q_Y \}$ 
to have a solution. In the simplest minimal case with $c_{n_Y>1}=0$ 
(and then $K_{Y\bar{Y}}=c_1>0$), a negative charge, $q_Y<0$,  
is required. 

A nontrivial point of this solution is that this SUSY 
stationary point is always tachyonic as we can see from the 
arguments in Appendix~\ref{app:rmass}. In addition, we can find a 
SUSY breaking minima along the direction $D_{\chi_i} \zeta=0$ 
(the first condition in Eq.~(\ref{eq:susyvc2})), 
if we assume that $\chi_i$ receives a heavy SUSY mass 
$m^2_{\chi_i} \gg m_{3/2}^2$ by the condition $D_{\chi_i} W=0$. 
This is a reasonable assumption because $\chi_i$ has a vanishing 
R-charge and $\zeta(\chi_i)$ in $W$ is assumed to be a generic function. 

The scalar potential along $D_{\chi_i} \zeta=0$ is found to be 
\begin{eqnarray}
v(Y) &=& V \Big|_{D_{\chi_i} f=0} \ = \ 
e^K \Big( K_{Y\bar{Y}}^{-1}|2/q_Y+K_Y Y|^2
-3|Y|^2 \Big) |Y|^{2(2/q_Y-1)}|\zeta|^2. 
\nonumber
\end{eqnarray}
Again, for the minimal K\"ahler potential (\ref{eq:kxpol}) 
with $c_1=1$ and $c_{n_Y>1}=0$, the stationary condition 
\begin{eqnarray}
\partial_Y v(Y) &=& 
e^{\hat{K}( \langle \chi_i \rangle, \langle \bar\chi_i \rangle)} 
e^{|Y|^2} |Y|^{2/q_Y-2} (2/q_Y+|Y|^2)
\nonumber \\ && \times 
\Big( |Y|^4+2(2/q_Y-1)|Y|^2+(2/q_Y)^2-2/q_Y \Big) 
\ = \ 0, 
\nonumber
\end{eqnarray}
leads to solutions 
\begin{eqnarray}
|Y|^2 &=& -2/q_Y, 
\label{eq:susysp}
\end{eqnarray}
and 
\begin{eqnarray}
|Y|^2 &=& 1-2/q_Y \pm \sqrt{1-2/q_Y}. 
\label{eq:susubm}
\end{eqnarray}
The first solution (\ref{eq:susysp}) corresponds to the 
SUSY saddle point and the second solutions 
(\ref{eq:susubm}) are SUSY breaking minima. 
We can find this kind of SUSY breaking minima 
in a similar way for more generic K\"ahler potential. 

We can study the same system in a different view point. 
We redefine the field $Y$ as 
\begin{eqnarray}
T &=& -\frac{2}{aq_Y} \ln Y, 
\label{eq:def:t}
\end{eqnarray}
where $a$ is a real constant. 
In this basis, the K\"ahler potential and the superpotential 
(\ref{eq:rsp1}) is written as 
\begin{eqnarray}
K &=& K(T+\bar{T},\chi_i,\bar\chi_i), 
\nonumber \\
W &=& e^{-aT} \zeta(\chi_i). 
\label{eq:rsp3}
\end{eqnarray}
This type of K\"ahler and superpotential appear in the 
four-dimensional effective theory derived from superstring theory, 
where $T$ may be a modulus field associated to some 
compactified dimensions. In such a case, the K\"ahler 
potential is typically given by 
\begin{eqnarray}
K &=& -n_T \ln (T+\bar{T})
+\hat{K}(\chi_i,\bar\chi_i), 
\nonumber
\end{eqnarray}
where $n_T$ is a fractional number, and the $T$-dependence 
of the superpotential (\ref{eq:rsp3}) may originate from 
nonperturbative effects such as string/D-brane instanton effects 
and gaugino condensation effects, where the corresponding gauge coupling 
is determined by the vacuum value of $T$. 
In this case, the scalar potential along $D_{\chi_i} \zeta=0$ 
is given by 
\begin{eqnarray}
v(T) &=& V \Big|_{D_{\chi_i} f=0} \ = \ 
e^K \Big( K_{T\bar{T}}^{-1}(K_T-a)^2-3 \Big) |e^{-aT} \zeta |^2, 
\nonumber
\end{eqnarray}
and then the stationary condition
\begin{eqnarray}
\partial_t v(t) &=& 
-e^{\hat{K}( \langle \chi_i \rangle, \langle \bar\chi_i \rangle)} 
e^{-at}t^{-n_T-1}
\nonumber \\ && \times 
(at+n_T)
\Big( (a^2/n_T) t^2
+2a(1-1/n_T)t+n_T-3 \Big) \ = \ 0, 
\nonumber
\end{eqnarray}
results in a SUSY saddle point $t=-n/a$ and 
SUSY breaking minima 
\begin{eqnarray}
t &=& -(n_T/a)(1-1/n_T) 
\pm (n_T|a|/a^2)\sqrt{5/n_T+1/n_T^2}, 
\nonumber
\end{eqnarray}
where $t=T+\bar{T}$.

In the literature, there are examples of the models 
which have this kind of vacuum structure of the potential. 
Typical superstring models have several moduli $T_I$ with the K\"ahler 
potential $K=\ln \prod_I (T_I+\bar{T}_I)^{-n_{T_I}}$. 
The superpotential induced by some nonperturbative effects 
is given by 
\begin{eqnarray}
W &=& \sum_n A_n e^{\sum_I a_n^I T_I}, 
\nonumber
\end{eqnarray}
where $A_n$ and $a_n^I$ are constants. If the number of the 
moduli is the same as or larger than the number of the 
nonperturbative terms appearing in the 
superpotential~\cite{Barreiro:1999hp}, 
we can define an R-symmetry. A particular linear combination of 
$T_I$'s corresponds to $T$ in Eq.~(\ref{eq:rsp3}) which is 
determined by the condition that all the remaining 
combinations corresponding $\chi_i$'s receive a heavy 
mass by the SUSY condition $D_{\chi_i}W=0$. 
This is possible for certain values of $a_n^I$. 
For the two moduli with double nonperturbative terms, i.e., 
racetrack models,  
a detailed analysis was carried out in Ref.~\cite{Abe:2005pi}.

We stress that the analysis of the SUSY breaking 
minimum as well as the SUSY saddle point in this 
subsection is based on the assumption that all the other fields 
$\chi_i$ than $Y$ or $T$ are stabilized by $D_{\chi_i}W=0$, 
that is, by the SUSY masses~\cite{deAlwis:2005tf}. 
We comment that these stationary solutions have a nonvanishing 
and negative vacuum energy. We need to uplift the SUSY 
breaking minimum to a Minkowski vacuum in order to identify 
this minimum as the one we are living. For such purpose, 
we need another sector which provides the uplifting energy 
and is well sequestered in order not to spoil the original 
structure of dynamical SUSY breaking. 
Such sector can be realized by a dynamically generated 
F-term~\cite{Lebedev:2006qq,Abe:2006xp} for which the discussions 
in the following sections would be important. 

In summary, there is a possibility of special SUSY 
stationary solution in R-symmetric supergravity with a 
generic superpotential. However, it is always a saddle point 
at best and we find SUSY breaking minima with lower 
vacuum energy. This may imply that the NS argument for a dynamical 
SUSY breaking is qualitatively correct also in this 
case, although there is a SUSY solution. 
Furthermore,  the NS argument 
still holds in supergravity as long as the 
K\"ahler potential satisfies $2/q_Y+YK_Y \ne 0$ for any value of $Y$ 
in the field basis (\ref{eq:rbasis1}). 
For instance, in typical models with $q_Y>0$ 
and $K=|Y|^2$, we always find $2/q_Y+YK_Y > 0$.

\subsection{Generalized O'Raifeartaigh model in supergravity}
Now we consider the generalized OR model (\ref{eq:orgsp}) in supergravity. 
The F-flat conditions (\ref{eq:orgfflat}) for $X_a$ become  
\begin{eqnarray}
D_{X_a} W &=& 
\partial_{X_a} W + (\partial_{X_a} K)\,W 
\nonumber \\ &=& 
\sum_b M_{ab}(X_c,\phi_i)\,(g_b(\phi_i) + \delta_{b1}f) 
\ = \ 0, 
\label{eq:orgfflatsugra}
\end{eqnarray}
where 
\begin{eqnarray}
M_{ab}(X_c,\phi_i) &=& 
\delta_{ab}+K_{X_a}X_b. 
\nonumber
\end{eqnarray}
We define its determinant as 
\begin{eqnarray}
\Delta &\equiv& \det M_{ab} 
\ = \ 1+\sum_a K_{X_a} X_a. 
\label{eq:det1kx}
\end{eqnarray}

If there is no solution for $\Delta=0$, the matrix $M_{ab}$ 
has an inverse matrix and consequently the F-flat conditions 
(\ref{eq:orgfflatsugra}) are reduced to the same ones as 
Eq.~(\ref{eq:orgfflat}) in the global SUSY, 
\begin{eqnarray}
g_a(\phi_i) + \delta_{a1}f&=& 0, 
\nonumber
\end{eqnarray}
which does not allow a solution for $r>s$ in general. 
However, in the limit $f \rightarrow 0$ in the tilde basis 
(\ref{eq:orgsprd}), these equations are satisfied at $\phi_i=0$.
Thus, the constant $f$ represents the typical size of 
SUSY breaking effects and $g_a(\phi_i)$ as the global 
SUSY case.
We comment that the situation changes if there exists a 
solution of $\Delta=0$. Actually, the condition $\Delta=0$ 
is an analogue of the second condition in Eq.~(\ref{eq:susyvc2}). 
Then, we can carry out a similar analysis as in 
the previous subsection also for this OR model.
That is straightforward and is 
omitted here. Note that the condition $\Delta=0$ is never satisfied 
for a minimal K\"ahler potential, 
\begin{eqnarray}
K &=& \sum_a |X_a|^2 +\sum_i|\phi_i|^2. 
\label{eq:minkahler}
\end{eqnarray}
In the following, we just assume that there is no solution 
for $\Delta=0$. 

We comment that, even in supergravity, the scalar potential 
is positive,  $V >  0$, in the generalized OR model 
(\ref{eq:orgsprd}) with the minimal K\"ahler potential 
(\ref{eq:minkahler}).
In this case, the scalar potential is written as 
\begin{eqnarray}
V &=& e^{K}\left[ (\bar{g}_a +\delta_{a1}\bar f) \big\{ \delta_{ab} 
+(|X_c|^2 -1) \bar{X}_a X_b \big\} (g_b + \delta_{b1}f)
+|X_a D_{\phi_i} g_a|^2  \right] .
\nonumber 
\end{eqnarray}
For any vacuum values of $X_a$, we can always rotate their basis as 
\begin{eqnarray}
& & U_{ab}\,X_b =\hat{X}_a 
=(0,\ldots,0,\hat{X}_c,0,\ldots,0),
\nonumber
\end{eqnarray}
 by a unitary matrix $U(X_a)$, 
and in this basis we can write 
\begin{eqnarray}
e^{-K}V &=& 
\big\{ (|\hat{X}_c|^2-1/2)^2+3/4 \big\} |\hat{g}_c|^2
+ \sum_{a \neq c} |\hat{g}_a|^2
+\sum_{i}|\hat{X}_c D_{\phi_i} \hat{g}_c|^2 
\ > \ 0, 
\nonumber
\end{eqnarray}
where $\hat{g}_a=(U^\dagger)_{ab}\,(g_b +\delta_{b1}f)$. 
Note that $\hat{g}_a$ are now 
$X_a$-dependent functions. 
As discussed above, the conditions, $\hat g_a(\phi)=0$,  
can not be satisfied at the same time.
Thus, the vacuum energy must be positive, $V > 0$.
Since typical magnitudes of $\hat g_a(\phi)$ would be 
of ${\cal O}(f)$, we would estimate $V \sim f^2$.
To realize the almost vanishing vacuum energy $V \approx 0$ 
at this SUSY breaking minimum, we need 
a negative and sizable contribution to 
the vacuum energy, which can be generated by 
R-symmetry breaking effects, e.g., the constant term 
in the superpotential.

\section{Explicit R-symmetry breaking in supergravity}
\label{sec:r-breaking-sugra}

Here we study explicit R-symmetry breaking terms 
in supergravity and 
examine whether SUSY solutions can be found
by adding explicit R-symmetry breaking terms 
to the NS model and the generalized OR model.
In the previous section, we have pointed out that 
there is a SUSY 
stationary point when the condition (\ref{eq:susyvc2:2}) 
or the condition $\Delta =0$ is satisfied.
In the following sections, we consider the models, where 
such conditions are not satisfied, and 
SUSY is broken in the NS and generalized OR models 
even within the framework of supergravity like global 
SUSY theory.

First we consider the NS model with 
explicit R-symmetry breaking terms 
$W_{R\!\!\!\!/}=w(Y,\chi_i)$. 
The total superpotential is written as,
\begin{eqnarray}
 & & W=Y^{2/q_Y}\zeta(\chi_i) + w(Y,\chi_i).
\nonumber 
\end{eqnarray}
In this case, F-flat conditions of supergravity theory, 
$D_YW=D_{\chi_i} W=0$, do not lead to {\it over-constrained} 
conditions for any non-vanishing function $w(Y,\chi_i)$.
It is remarkable that within the framework of 
supergravity theory the constant term 
$W_{R\!\!\!\!/} = c$ breaks 
R-symmetry and even such term is enough to 
relax the  {\it over-constrained} conditions.

\subsection{Generalized O'Raifeartaigh model}
Let us study more explicitly the generalized OR model 
with explicit R-symmetry breaking terms 
$W_{R\!\!\!\!/}=w(X_a,\phi_i)$.
The total superpotential is written as,
\begin{eqnarray}
 & & W=fX_1 +\sum_{a=1}^rg_a(\phi_i)X_a+w(X_a,\phi_i).
\nonumber
\end{eqnarray}
First, we consider the case with 
the constant R-symmetry breaking term, 
$W_{R\!\!\!\!/}=c$.
In this case, F-flat conditions are 
written explicitly as 
\begin{eqnarray}
D_{X_a}W &=& f\delta_{a1}+g_a(\phi_i) + K_{X_a}\left( 
fX_1 +\sum_{a=1}^rg_a(\phi_i)X_a+c \right) \ = \ 0, 
\nonumber \\
D_{\phi_i}W &=& \sum_a X_a \partial_i g_a(\phi_i) 
+ K_{\phi_i}\left( 
fX_1 +\sum_{a=1}^rg_a(\phi_i)X_a+c \right) \ = \ 0.
\nonumber 
\end{eqnarray}
The former conditions are not always {\it over-constrained} 
for $c \neq 0$.
Furthermore, the vacuum expectation value of $W$ and 
at least $(r-s)$ vacuum values of $K_{X_a}$ are required to be 
non-vanishing.
Otherwise, the former conditions become {\it over-constrained}  
for generic functions $g_a(\phi_i)$.
Furthermore, when  $K_{X_a}$ for $a\neq 1$ does 
not vanish, all vacuum values of $\phi_i$ can not 
vanish to satisfy $D_{X_a}W=g_a(\phi_i)+K_{X_a}W=0$.
Thus, a SUSY solution can be found by 
adding $W_{R\!\!\!\!/}=c$.
This solution corresponds to 
the AdS SUSY minimum, because 
non-vanishing $\langle W \rangle$ is required 
and the scalar potential at this point 
is evaluated as $V=-3e^K|W|^2 < 0$.
The values of the constant $c$ and $\langle W \rangle$ 
must be sizable, because this AdS SUSY point disappears 
in the limit that $c\rightarrow 0$ or 
$ \langle W \rangle \rightarrow 0$.
Magnitudes of  $c$ and $\langle W \rangle$ are 
expected to be comparable with $f$ when 
$K_{X_a}={\cal O}(1)$.
Hence, we can find the new type of 
SUSY solution, which can not be found 
in global SUSY theory. 
However, that requires large  values of $c$ and $\langle W \rangle$, 
 and may have sizable effects on the previous SUSY breaking 
minimum, which is found in the generalized OR model without 
R-symmetry breaking terms.

Similarly, we can discuss the case that 
R-symmetry breaking terms include only $\phi_i$ 
fields, i.e.,  $W_{R\!\!\!\!/}=w(\phi_i)$.
In this case, F-flat conditions along $X_a$, $D_{X_a}W=0 $, 
are written as 
\begin{eqnarray}
D_{X_a}W &=& f\delta_{a1}+g_a(\phi_i) + K_{X_a}\left( 
fX_1 +\sum_{a=1}^rg_a(\phi_i)X_a+ w(\phi_i)\right) \ = \ 0 .
\nonumber 
\end{eqnarray}
Thus, the situation is quite similar to the case 
with $W_{R\!\!\!\!/}=c$.
To have a SUSY solution, it is required that 
$\langle W \rangle$, $\langle w(\phi_i) \rangle $ 
and at least $(r-s)$ vacuum values of $K_{X_a}$ must 
be non-vanishing.
Sizes of $\langle W \rangle$ and $\langle w(\phi_i) \rangle$ 
are expected to be comparable with $f$.

Finally, we consider the case that R-symmetry breaking terms include 
$X_a$ fields, $W_{R\!\!\!\!/}=w(X_a,\phi_i)$.
In this case, F-flat conditions along $X_a$, $D_{X_a}W=0 $, 
are written as 
\begin{eqnarray}
D_{X_a}W &=& f\delta_{a1}+g_a(\phi_i) + \partial_{X_a}w(X_a,\phi_i)
+ K_{X_a}W =0.
\nonumber 
\end{eqnarray}
When $K_{X_a}W$ is sufficiently small, 
the above F-flat conditions correspond to 
F-flat conditions in global SUSY theory.
In such a case, we have a SUSY solution when 
$w(X_a,\phi_i)$ depend on at least $(r-s)$ $X_a$'s.
Otherwise, the global SUSY solution can not 
be found, but a SUSY solution with 
$\langle w(X_a,\phi_i) \rangle \neq 0$ 
and $\langle W \rangle \neq 0$ can be found 
within the framework of supergravity theory.
Such situation is similar to the case with $W_{R\!\!\!\!/}=c$.

We have discussed that the NS model and generalized OR 
model with R-symmetry breaking terms have 
SUSY solutions with $\langle W \rangle \neq 0$ 
in supergravity theory.
If the SUSY breaking minimum, which is found 
without R-symmetry breaking terms, is not destabilized 
by the presence of R-symmetry breaking terms, 
the previous SUSY breaking minimum 
would correspond to a SUSY breaking metastable vacuum.
However, a sizable vacuum value of superpotential 
is required unless $\partial_{X_a}w(X_a,\phi_i) \neq 0$ 
for at least $(r-s)$ $X_a$ fields.
Such large superpotential (even if that is a constant term) 
would affect the stability of 
the previous SUSY breaking minimum.

Furthermore, we have another reason to have 
a large size of $\langle W \rangle$ at 
the previous SUSY breaking minimum.
At the previous SUSY breaking minimum, the vacuum 
energy is estimated as 
$V \sim |f|^2 > 0$ for $\langle W \rangle =0$.
To realize the almost vanishing vacuum energy, $V\approx 0$, 
we need a non-vanishing value of $\langle W \rangle$, 
which are comparable with $f$.
In this case, supergravity effects at the 
previous SUSY breaking minimum are not negligible.
This purpose to realize  $V\approx 0$ has 
the implication even for the case that 
R-symmetry breaking terms include more than $(r-s)$ 
$X_a$ fields.
In this case, we can find a (global) SUSY solution even for 
$\langle W \rangle =0$.
However, realization of $V\approx 0$ requires 
a sizable vacuum value of $\langle W \rangle $, 
although values $\langle W \rangle $ at the 
SUSY breaking minimum and SUSY preserving minimum 
are not the same.
Hence, it is quite non-trivial whether 
one can realize a metastable SUSY breaking vacuum 
with  $V\approx 0$ in supergravity theory, 
which has a SUSY minimum.
We will study this possibility concretely  
by using simple classes of the generalized OR models 
in the following sections.
We will concentrate ourselves to the minimal K\"ahler potential 
(\ref{eq:minkahler}) in most cases of the following 
discussions.

\subsection{Classification of R-breaking terms in supergravity}

In this subsection and the following sections, 
we consider minutely the previous discussions 
about the explicit R-symmetry breaking in the supergravity 
framework by examining concrete examples. 
We introduce the explicit R-symmetry breaking terms 
$W_{R\!\!\!\!/}$ into the above supergravity OR model, 
\begin{eqnarray}
W_{R\!\!\!\!/} &=& c(\phi_i)
+\frac{1}{2}\sum_{a,b}
m\,\epsilon_{ab}(\phi_i) X_a X_b 
+\cdots, 
\label{eq:rbrkexg}
\end{eqnarray}
where $c(\phi_i)$ and $\epsilon_{ab}(\phi_i)$ 
are generic functions of $\phi_i$ including $\phi$-independent 
constants, and the ellipsis denotes the higher order terms in $X_a$. 
Note that, as mentioned before, only the $\epsilon_{ab}(\phi_i)$ terms 
are relevant to the recovery of SUSY vacua in the case of global SUSY. 
Now we have the total superpotential, $W=W_{OR}+W_{R\!\!\!\!/}$.
The F-flat conditions (\ref{eq:orgfflatsugrard}) are modified as 
\begin{eqnarray}
D_{X_a}W &=& 
\sum_b M_{ab} \bigg( g_b(\phi_i) +\delta_{a1}f_1 
+\sum_{c,d}  M^{-1}_{bc} \epsilon_{cd}(\phi_i) X_d 
+\Delta^{-1} K_{X_b} W_{R\!\!\!\!/} \bigg) \ = \ 0. 
\label{eq:orgfflatsugrarb}
\end{eqnarray}
Here we find that all the terms in $W_{R\!\!\!\!/}$ 
including $c(\phi_i)$ are accompanied by $X_a$ in the 
above F-flat conditions and then have a possibility 
for restoring SUSY, contrary to the case of 
global SUSY explained in the previous section. 

Most notably, just a constant superpotential 
\begin{eqnarray}
W_{R\!\!\!\!/} &=& c, 
\label{eq:constantsp}
\end{eqnarray}
i.e., $c(\phi_i)=c$ and $\epsilon_{ab}(\phi_i)=0$, can restore 
SUSY. In this case with the minimal K\"ahler potential 
(\ref{eq:minkahler}), we find a solution for 
Eq.~(\ref{eq:orgfflatsugrarb}) as 
\begin{eqnarray}
\bar{X}_a &=& -c^{-1} \Delta\,g_a(\phi_i), 
\label{eq:orgsugragsol}
\end{eqnarray}
where $\Delta=1+\sum_a|X_a|^2$ defined in Eq.~(\ref{eq:det1kx}) 
is real and positive. 
{}From Eq.~(\ref{eq:orgsugragsol}), $X_a$ can be written in terms 
of $\phi_i$, and then $\Delta$ is given by 
\begin{eqnarray}
\Delta &=& \frac{|c|^2 \pm |c| 
\sqrt{|c|^2-4 \sum_a |g_a(\phi_i)|}}{2 \sum_a |g_a(\phi_i)|}, 
\nonumber
\end{eqnarray}
which should be a real number. 
Therefore, in order for the SUSY solution (\ref{eq:orgsugragsol}) to be 
valid, the constant superpotential $c$ must satisfy the condition 
\begin{eqnarray}
4 \sum_a |g_a( \langle \phi_i \rangle )|^2 \le |c|^2, 
\label{eq:orgsugrassc}
\end{eqnarray}
where $\langle \phi_i \rangle$ are solutions of $D_{\phi_i}W=0$ 
under the condition (\ref{eq:orgsugragsol}). 

Because $X_1$ is distinguished in the superpotential 
(\ref{eq:orgsprd}), we divide the generic R-breaking terms 
(\ref{eq:rbrkexg}) into two pieces: 
\begin{eqnarray}
W_{R\!\!\!\!/} &=& 
W^{(A)}_{R\!\!\!\!/}+W^{(B)}_{R\!\!\!\!/}, 
\nonumber
\end{eqnarray}
where 
\begin{eqnarray}
W^{(A)}_{R\!\!\!\!/}(X_{a \ne 1};\,\phi_i) &=& 
c(\phi_i)+\frac{1}{2}\sum_{a,b \ne 1}
m\,\epsilon_{ab}(\phi_i) X_a X_b 
+\cdots, 
\label{eq:arb} \\
W^{(B)}_{R\!\!\!\!/}(X_1;\,X_{a \ne 1},\phi_i) &=& 
\sum_{a \ne 1} m\,\epsilon_{a1}(\phi_i) X_a X_1 
+\frac{1}{2} m\,\epsilon_{11}(\phi_i) X_1^2 
+\cdots. 
\label{eq:brb}
\end{eqnarray}
The ellipses denote the higher order terms 
in terms of $X_{a \ne 1}$ in $W^{(A)}_{R\!\!\!\!/}$, 
and those of $X_1$ and $X_{a \ne 1}$ in $W^{(B)}_{R\!\!\!\!/}$. 
Without loss of generality, we can assume that $\epsilon_{11}(0)$ 
is real and positive among $\epsilon_{ab}(0)$, which is referred 
as $\epsilon$ in Sec.~\ref{sec:typeb}.

\section{Type-A breaking: Polonyi-like models}
\label{sec:typea}
In this section, we study the effect of R-breaking terms 
(\ref{eq:arb}) which we call the A-type breaking, 
$$
W \ = \ W_{OR}+W^{(A)}_{R\!\!\!\!/}. 
$$ 
Because this type of breaking terms does not contain $X_1$, 
we find the Polonyi model~\cite{Polonyi:1977pj} 
\begin{eqnarray}
W \big|_{X_{a \ne 1}=0,\phi_i=0} 
&=& W_{\rm Polonyi} \ \equiv \ fX_1+c, 
\label{eq:polonyi}
\end{eqnarray}
in the hypersurface $X_{a \ne 1}=0$, $\phi_i=0$ of the 
scalar potential, where $c=c(0)$. 
This hypersurface would be a stationary 
plane in the $X_{a \ne 1}$- and the $\phi_i$-directions 
if $\partial_{\phi_i} g_{a \ne 1}(0)$ are sufficiently 
large, which correspond to SUSY masses for 
$X_{a \ne 1}$ and $\phi_i$ on that plane. 

Moreover, if $m_1^i$ and/or $h_1^{ij}$ in 
\begin{eqnarray}
g_1(\phi_i) &=& m_1^i\phi_i +h_1^{ij} \phi_i \phi_j +\cdots, 
\label{eq:hx1phi2}
\end{eqnarray}
are nonvanishing, the Polonyi model in this hypersurface 
can be affected/modified by a tree-level SUSY mass 
and/or a one-loop SUSY breaking mass for $X_1$. 
Then, we further classify the A-type breaking models into 
two cases, $g_1(\phi_i)=0$ and $g_1(\phi_i) \ne 0$.

\subsection{Decoupled case: $g_1(\phi_i)=0$}
In the case with $g_1(\phi_i)=0$, the superpotential of 
the A-type breaking models is written as 
\begin{eqnarray}
W &=& fX_1 +\sum_{a \ne 1} g_a(\phi_i)X_a 
+c(\phi_i)
+\frac{1}{2} \sum_{a,b \ne 1}m 
\epsilon_{ab}(\phi_i) X_a X_b +\cdots 
\nonumber \\ &=& 
c+fX_1+\frac{1}{2}\mu_{AB}\Phi_A \Phi_B +\cdots, 
\nonumber
\end{eqnarray}
where $\Phi_A=(X_{a \ne 1},\phi_i)$ with the index $A=(a \ne 1,i)$. 
The SUSY mass matrix $\mu_{AB}$ is given by 
the R-breaking components, 
$\mu_{a \ne 1,b \ne 1}=m\epsilon_{ab}(0)$, 
$\mu_{ij}= \partial_{\phi_i} \partial_{\phi_j} c(0)$ and 
the R-symmetric components, 
$\mu_{a \ne 1,i}=2 \partial_{\phi_i} g_a(0)$. 
After the unitary rotation which makes $\mu_{AB}$ diagonal, 
the above superpotential takes the form of 
\begin{eqnarray}
W &=& 
c+fX_1+\frac{1}{2}\mu_{A} \Phi_A^2 +\cdots, 
\label{eq:modelad}
\end{eqnarray}
where $\mu_A$ represents the eigenvalues of $\mu_{AB}$. 
Because of the SUSY mass $\mu_A$, the field 
$\Phi_A$ would be integrated out without affecting the 
low energy dynamics of $X_1$, because $X_1$ is completely 
decoupled in the present case\footnote{We may have to 
assume that the K\"ahler mixing is also zero or negligible 
between $X_1$ and the others.}. 

Then, the effective action for $X \equiv X_1$ is just determined 
by the Polonyi superpotential (\ref{eq:polonyi}), where the 
phase of $c$ and $f$ can be eliminated by the $U(1)_R$ 
rotation and the rephasing of $X_1$. Assuming the minimal 
K\"ahler potential (\ref{eq:minkahler}) for simplicity, 
the effective scalar potential is minimized by a real vacuum 
value $X=\bar{X}=x$ satisfying the stationary condition 
\begin{eqnarray}
V_X &=& e^GG_X(G_{XX}+G_X^2-2) \ = \ 0, 
\nonumber
\end{eqnarray}
where $G=K+\ln |W|^2$ and
\begin{eqnarray}
G_{XX}+G_X^2-2 &=& fW^{-1}(x^3 +f^{-1}c x^2 -2f^{-1}c), 
\label{eq:vxsb} \\
G_X &=& fW^{-1}(x^2 +f^{-1}c x +1). 
\label{eq:vxss}
\end{eqnarray}
The F-flat condition for $X$ corresponds to $G_X=0$, and 
the SUSY breaking stationary point is determined 
by the condition $G_{XX}+G_X^2-2=0$. 

As we declared, we persist in obtaining a vanishing vacuum 
energy at the SUSY breaking minimum. Then in 
addition to the stationary condition $G_{XX}+G_X^2-2=0$, 
we set $V=e^G(G^{X \bar{X}}|G_X|^2-3)=0$. 
In this case, we have to take a definite value 
of the constant $c$ and find two solutions 
\begin{eqnarray}
(x,\,f^{-1}c) &=& 
(\sqrt{3}-1,\,2-\sqrt{3}), 
\label{eq:plsol1}
\end{eqnarray}
and 
\begin{eqnarray}
(x,\,f^{-1}c) &=& 
(-\sqrt{3}-1,\,2+\sqrt{3}). 
\label{eq:plsol2}
\end{eqnarray}
The mass eigenvalues of $({\rm Re}X,\,{\rm Im}X)$ are computed 
as $(2\sqrt{3}f^2,\,(4-2\sqrt{3})f^2)$ for the first solution 
(\ref{eq:plsol1}), and $(-2\sqrt{3}f^2,\,(4+2\sqrt{3})f^2)$ for 
the second one (\ref{eq:plsol2}), at this SUSY breaking 
Minkowski stationary point where $W=f$. Then, only the first 
solution (\ref{eq:plsol1}) can be a minimum of the potential, 
while the second one (\ref{eq:plsol2}) is a saddle point. 
We comment that $\phi_i$ and $X_{a \ne 1}$ directions would not 
possess tachyonic masses at these points for sufficiently large 
SUSY mass $\mu_{A}$ compared with the SUSY 
breaking mass $f$. Therefore, the candidate for our present universe, 
where the SUSY is broken with (almost) vanishing vacuum energy, 
is the first solution (\ref{eq:plsol1}).

In addition to a SUSY breaking solution satisfying 
$G_{XX}+G_X^2-2=0$, we have a SUSY solution 
$G_X=0$ due to the R-breaking effect $c \ne 0$, that is, 
\begin{eqnarray}
x_\pm &=& \frac{1}{2}(-f^{-1}c \pm \sqrt{(f^{-1}c)^2-4}), 
\label{eq:plsusy}
\end{eqnarray}
if the R-breaking constant $c$ satisfies 
\begin{eqnarray}
|f^{-1}c| \ge 2. 
\label{eq:plsusycd}
\end{eqnarray}
Note that this condition (\ref{eq:plsusycd}) corresponds to 
Eq.~(\ref{eq:orgsugrassc}) in the previous general argument 
for the generalized OR model. 
The mass eigenvalues of $({\rm Re}X,\,{\rm Im}X)$ are computed as 
$W_\pm^2 (x_\pm^2-2)(x_\pm^2+1)$ and $W_\pm^2 (x_\pm^2-1)(x_\pm^2+2)$ 
at this SUSY AdS stationary point where 
\begin{eqnarray}
|W_\pm| &=& |fx_\pm +c| 
\ = \ \frac{1}{2}
\left| f(f^{-1}c \pm \sqrt{(f^{-1}c)^2-4}) \right| \ > \ 0, 
\nonumber
\end{eqnarray}
and then we obtain 
\begin{eqnarray}
V &=& -3e^G \ = \ -3e^{x_\pm^2}|W_\pm|^2 \ < \ 0. 
\nonumber
\end{eqnarray}
Remark that, in the vanishing (one of) R-breaking limit, 
$c \to 0$, the condition (\ref{eq:plsusycd}) is not satisfied, 
and the SUSY solution (\ref{eq:plsusy}) disappears. 
In the other words, this SUSY solution is a 
consequence of the R-breaking constant term $c$ in the 
superpotential. Due to the appearance of this SUSY 
solution, there is a possibility that the SUSY breaking 
point determined by $G_{XX}+G_X^2-2=0$ becomes a metastable vacuum 
as in the case of global SUSY explained previously. 

However, this is not the case. Interestingly, if we tune the 
R-breaking constant superpotential $c$ as $f^{-1}c=2-\sqrt{3}$ 
so that the solution (\ref{eq:plsol1}) with the vanishing vacuum 
energy is realized, the condition 
(\ref{eq:plsusycd}) is not satisfied and the SUSY 
stationary solution (\ref{eq:plsusy}) disappears. In such a sense, 
the constant R-breaking term $c$ does not lead to a metastability 
of SUSY breaking Minkowski minimum (\ref{eq:plsol1}).

Next, we consider the SUSY stationary solutions 
outside the Polonyi slice $X_{a \ne 1}=0$, $\phi_i=0$. 
For the superpotential (\ref{eq:modelad}), 
the F-flat directions are determined by 
\begin{eqnarray}
D_{\Phi_A}W &=& K_A W +\mu_A \Phi_A +\cdots \ = \ 0, 
\nonumber \\
D_{X_1}W &=& f +K_{X_1}W \ = \ 0, 
\nonumber
\end{eqnarray}
which can be satisfied by distinguishing a single field 
$\Phi_B \ne 0$ for  $^\exists B$ as 
\begin{eqnarray}
W &=& -K_B^{-1} \Phi_B (\mu_B +\cdots) 
\ = \ -K_{X_1}^{-1} f \qquad (\textrm{for} \ ^\exists B), 
\label{eq:ospsusysol} \\
\Phi_A &=& K_A \ = \ 0 \qquad 
(\textrm{for} \ A \ne B), 
\nonumber
\end{eqnarray}
where the ellipsis represents the higher order terms of $\Phi_B$. 
The first line gives two complex equations 
for two complex variables $X_1$ and $\Phi_{^\exists B}$, 
which have a solution in general. 

For example, if the K\"ahler potential is minimal 
(\ref{eq:minkahler}), all the parameters in the 
superpotential are real and there is no higher order terms 
of $\Phi_B$ (no ellipses in the above expressions), 
then the solution for Eq.~(\ref{eq:ospsusysol}) is found as 
\begin{eqnarray}
|\Phi_B|^2 &=& 
-2 \left( \frac{c}{\mu_B}
+\frac{f^2}{\mu_B^2} +1 \right) \ > \ 0, \qquad 
\Phi_{A \ne B} \ = \ 0. 
\label{eq:newsol}
\end{eqnarray}
For this value of $\Phi_B$, the remaining condition 
$D_{X_1}W=0$ is satisfied by 
$$
X_1=f/\mu_B. 
$$
Note that the number of this SUSY points 
is $n_X+n_\phi-1$ because the solution (\ref{eq:newsol}) 
is valid for every choice of $B=(b \ne 1,j)$. 
In order for the solution (\ref{eq:newsol}) to be valid, 
the parameter $\mu_B$ must satisfy 
\begin{eqnarray}
\mu_B^2 +c\mu_B +f^2 \ \le \ 0. 
\nonumber
\end{eqnarray}
This leads to the same condition (\ref{eq:plsusycd}) 
for the R-breaking constant term $c$ as in the 
Polonyi-type SUSY solution. 

In summary, the A-type breaking terms (\ref{eq:arb}) 
can restore SUSY in the generalized OR model (\ref{eq:orgsp}) 
or equivalently (\ref{eq:orgsprd}) in general. 
However, if we tune the R-breaking constant term in the 
superpotential so that the SUSY breaking minimum has a 
vanishing vacuum energy, i.e., (\ref{eq:plsol1}), the SUSY 
solutions (\ref{eq:plsusy}) and (\ref{eq:newsol}) disappear. 
Therefore, in this sense, the A-type R-symmetry breaking terms 
do not lead to a metastability of the SUSY breaking (Minkowski) 
vacuum aside from a possibility of the existence of more complicated 
SUSY solutions than (\ref{eq:newsol}).

\subsection{Generic case: $g_1(\phi_i) \ne 0$}
\label{sec:atypegen}
Now we turn on a nonvanishing $g_1(\phi_i)$ as in 
Eq.~(\ref{eq:hx1phi2}). With this term, the tree-level 
(field dependent) mass matrices in the $\phi_i=0$ plane 
contain the following contributions, 
\begin{eqnarray}
V_{X_1 \bar{X_1}} \big|_{\phi_l=0} 
&=& |m_1^i|^2+\cdots, 
\nonumber \\
V_{\phi_i \bar\phi_{\bar{j}}} \big|_{\phi_l=0}
&=& m_1^i \bar{m}_1^{\bar{j}}
+4h_1^{ik} \bar{h}_1^{\bar{j}\bar{k}}|X_1|^2 +\cdots, 
\nonumber \\
V_{\phi_i \phi_j} \big|_{\phi_l=0} 
&=& h_1^{ij}\bar{f} +\cdots, 
\nonumber \\
V_{X_1 \phi_i} \big|_{\phi_l=0}
&=& 2h_1^{ij}\bar{m}_1^{\bar{j}}X_1 +\cdots, 
\label{eq:g1mass}
\end{eqnarray}
where the ellipses represent the original terms 
involving $X_{a \ne 1}$, those coming from $c(\phi_i)$, 
and the supergravity corrections. Here the doubled 
indices are summed up. 
The K\"ahler covariant derivatives of the superpotential 
in the hypersurface $\phi_i=0$, $X_{a \ne 1}=0$ are given by 
$$
D_{X_1}W \big|_{\phi_l=0} \ = \ f +K_{X_1}W, \qquad 
D_{X_{a \ne 0}}W \big|_{\phi_l=0} \ = \ 0, \qquad 
D_{\phi_i}W \big|_{\phi_l=0} \ = \ m_1^i X_1. 
$$
{}From the third equation, we find that $\phi_i$ can not 
be integrated out prior to $X_1$ by the F-flat condition 
$D_{\phi_i}W=0$ unlike before. This is because, with the 
nonvanishing $m_1^i$, the source field $X_1$ for SUSY 
breaking shares a common SUSY mass with $\phi_i$ 
as shown in Eq.~(\ref{eq:g1mass}). 

In this case, the purely $X_1$-direction is no longer special 
in the scalar potential. We have to treat $X_1$ and $\phi_i$ 
at the same time. The analysis is quite complicated, and then 
we consider the case with $m_1^i=0$ in the following, where 
$g_1(\phi_i)$ starts from the quadratic term in $\phi_i$, and 
the $\phi_i$ can be integrated by their F-flat conditions 
$D_{\phi_i}W=0$ resulting $\phi_i=0$. We will comment about 
the case with $m_1^i \ne 0$ in Sec.~\ref{ssec:typebgen} together 
with more general R-breaking terms. 
The components of the mass matrices (\ref{eq:g1mass}) are 
now reduced to 
$$
V_{\phi_i \bar\phi_{\bar{j}}} \big|_{\phi_l=0}
\ = \ 
4h_1^{ik} \bar{h}_1^{\bar{j}\bar{k}}|X_1|^2 +\cdots, 
\qquad 
V_{\phi_i \phi_j} \big|_{\phi_l=0}
\ = \ 
h_1^{ij}\bar{f} +\cdots. 
$$
{}From the second equation, we observe that some linear 
combinations of ${\rm Re}\,\phi_i$ and ${\rm Im}\,\phi_j$ 
become tachyonic in the $\phi_i=0$ plane if $|h_1^{ij}f|$ 
dominate the SUSY mass for $\phi_i$. 
The $X_1$-dependence in the first one indicates that a 
SUSY breaking mass of $X_1$ is generated at the 
one-loop level, which is proportional to $h_1^{ij}$. 

Therefore, the effective potential after integrating out 
$\phi_i$ and $X_{a \ne 1}$ is given by 
\begin{eqnarray}
V &=& V^{(0)}+V^{(1)}, \qquad 
V^{(0)} \ = \ e^G (G^{X \bar{X}}|G_X|^2-3), \qquad 
V^{(1)} \ = \ m_X^2 |X|^2, 
\label{eq:qcorsugra}
\end{eqnarray}
where $X \equiv X_1$, $G=K+ \ln |W|^2$, and the effective 
superpotential $W=W_{\rm Polonyi}$ is shown in 
Eq.~(\ref{eq:polonyi}). The one-loop mass $m_X$ is 
determined by $h_1^{ij}$ as well as $f$, which would be 
considered as an independent parameter in the effective action. 
The stationary condition $V_X=0$ results in~\cite{Abe:2006xp} 
$$
X \ \simeq \ 2 fc/m_X^2, 
$$
for $c \sim f \sim m_X \ll 1$ in the unit with $M_{Pl}=1$, and 
the vanishing vacuum energy at this minimum requires 
$$
c \ = \ f/\sqrt{3} +{\cal O}(f^3/m_X^2). 
$$
The SUSY is broken at this Minkowski minimum 
with $D_XW=f+{\cal O}(f^2)$ and $W=f/\sqrt{3}+{\cal O}(f^2)$.

\section{Adding type-B breaking: Metastable universe}
\label{sec:typeb}
In the previous section, we have analyzed the generalized OR model 
with the explicit R-symmetry breaking terms (\ref{eq:brb}) 
which do not involve the source field $X_1$ for the 
dynamical SUSY breaking. 

In this section, we study more general case with the 
R-breaking terms (\ref{eq:arb}) including $X_1$, i.e.,  
$$
W \ = \ W_{OR}+W^{(A)}_{R\!\!\!\!/}+W^{(B)}_{R\!\!\!\!/}.
$$
In the type-B breaking terms (\ref{eq:arb}), the first term 
with $\epsilon_{a \ne 1,1}(0)$ gives the common SUSY 
mass for $X_1$ and $X_{a \ne 1}$ in the $\phi_i=0$ plane. 
Then the situation is similar to the case with a nonvanishing 
$m_1^i$ in Eq.~(\ref{eq:hx1phi2}), that is, we can not integrate 
out $X_{a \ne 1}$ prior to $X_1$, and we will include this case 
also in Sec.~\ref{ssec:typebgen}. 

By setting $\epsilon_{a \ne 1,1}(0)=0$, the superpotential 
in the hypersurface $\phi_i=X_{a \ne 1}=0$ is given by 
\begin{eqnarray}
W &=& fX+\frac{1}{2}m \epsilon X^2+c
+\cdots, 
\label{eq:qdcpl}
\end{eqnarray}
where $X \equiv X_1$, $\epsilon=\epsilon_{11}(0)$ 
and the ellipsis stands for the higher order terms in $X$.

\subsection{Decoupled case: $g_1(\phi_i) =0$}
As in the previous section, we first consider the case 
with $g_1(\phi_i) =0$, where $X_1$ is decoupled from the others in 
the superpotential. In this case the hypersurface $\phi_i=X_{a \ne 1}=0$  
would be stable in the $\phi_i$-, $X_{a \ne 1}$-direction 
as in Sec.~5.1. The effective theory in this slice is 
described by the superpotential (\ref{eq:qdcpl}). 

With the minimal K\"ahler potential (\ref{eq:minkahler}), 
real parameters $f$, $c$, $m$ and no higher order terms (ellipsis) 
in the superpotential (\ref{eq:qdcpl}) for simplicity, 
the SUSY breaking and SUSY stationary 
conditions are respectively given by Eqs.~(\ref{eq:vxsb}) and 
(\ref{eq:vxss}). In the limit $\epsilon \to 0$ of Eq.~(\ref{eq:qdcpl}), 
the SUSY breaking solution is given by Eq.~(\ref{eq:plsol1}). 
Then we can find the deviation of $X$ from this point assuming 
$\epsilon \ll 1$ and $m \sim c^{1/3} \sim f^{1/2}$. We find a 
SUSY breaking minimum with a vanishing vacuum energy at 
\begin{eqnarray}
X_{SB} &=& X_0 +\delta X, \qquad 
X_0 \ = \ \sqrt{3}-1, \qquad 
\delta X \ = \ 
-\frac{\epsilon m}{2f} +{\cal O}(\epsilon^2), 
\label{eq:btypesbmin}
\end{eqnarray}
where the constant superpotential term $c$ is tuned as 
\begin{eqnarray}
c &=& 
(2-\sqrt{3})f 
+(2 \sqrt{3}-3) \epsilon m 
+{\cal O}(\epsilon^2).
\label{eq:cbtype}
\end{eqnarray}

On the other hand, a SUSY solution, 
\begin{eqnarray}
X_{SUSY} &\simeq& -\frac{2f}{\epsilon m}, 
\label{eq:btypessmin}
\end{eqnarray}
arises as a consequence of the B-type R-breaking term represented by 
the parameter $\epsilon$, although the vacuum energy is set to be 
vanishing at the SUSY breaking minimum. This is unlike the 
case of SUSY solutions (\ref{eq:plsusy}) and (\ref{eq:newsol}) 
caused by the introduction of A-type R-breaking terms (\ref{eq:arb}). 
The shift of SUSY breaking minimum $\delta X$ in Eq.~(\ref{eq:btypesbmin}) 
is rewritten as 
$$
\delta X/X_0 \ \simeq \ 
\frac{1}{\sqrt{3}-1}\,\frac{1}{X_{SUSY}},
$$ 
and we find 
$$
|X_{SUSY}| \ > \ \frac{1}{\sqrt{3}-1} \ \sim \ {\cal O}(1),
$$
in order for the shift $\delta X$ to reside in a perturbative region, 
$|\delta X/X_0| < 1$. 

This means that the vacuum value of $|X|$ at the 
newly appeared SUSY vacuum must be larger than the Planck 
scale $M_{Pl}=1$, where the supergravity calculation might not valid. 
It would be possible that the potential is lifted for $|X|>1$ by 
the effect of quantum gravity, the above SUSY vacuum is 
washed out and the SUSY breaking minimum remains as a global 
minimum. If the supergravity approximation is valid even for $|X|>1$ 
by any reason, we obtain a constraint on the R-breaking parameter 
$\epsilon$ as 
$$\epsilon \ < \ 2(\sqrt{3}-1)|f/m|,$$ 
from the above condition. 

\begin{figure}[t]
\centerline{\epsfig{file=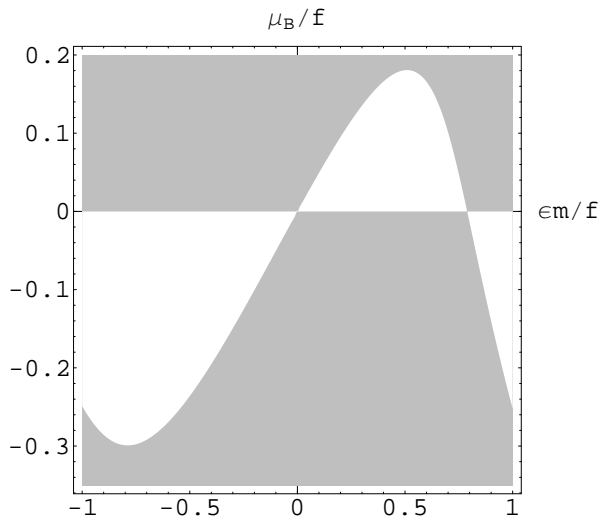,width=0.5\linewidth}}
\caption{Parameter region (white) of $\mu_B$ and $\epsilon$
allowing the SUSY solution (\ref{eq:newsolgen}). 
All the parameters are assumed to be real and the constant 
term $c$ is fixed by the vanishing vacuum energy condition 
(\ref{eq:cbtype}) at the SUSY breaking minimum 
(\ref{eq:btypesbmin}). In the shaded region, the SUSY 
solution (\ref{eq:newsolgen}) is not allowed and the 
SUSY breaking solution (\ref{eq:btypesbmin}) 
does not become metastable due to the R-breaking effect 
parameterized by $\epsilon$. We find no allowed region 
in the limit $\epsilon \to 0$ which corresponds to the 
solution (\ref{eq:newsol}).}
\label{fig:1}
\end{figure}

We also find a SUSY minimum outside the hyperplane 
$\phi_i=X_{a \ne 1}=0$, which is a generalization 
of Eq.~(\ref{eq:newsol}), given by 
\begin{eqnarray}
|\Phi_B|^2 &=& 
-\frac{2}{\mu_B}\, 
\left\{ \mu_B +c 
+\frac{f^2}{\mu_B-\epsilon m} 
\left( 1+\frac{\epsilon m}{2(\mu_B-\epsilon m)} \right) 
\right\} \ \ge \ 0, 
\nonumber \\
\Phi_{A \ne B} &=& K_{A \ne B} \ = \ 0, \qquad 
X \ = \ \frac{f}{\mu_B-\epsilon m}, 
\label{eq:newsolgen}
\end{eqnarray}
where we assumed the minimal K\"ahler potential 
(\ref{eq:minkahler}), and the absence of the higher 
order terms of $X$ in the superpotential for concreteness. 
In the limit $\epsilon \to 0$, this solution is 
reduced to (\ref{eq:newsol}). 
In contrast to (\ref{eq:newsol}), the above solution 
(\ref{eq:newsolgen}) does not disappear in all of the parameter 
region, even after the vacuum energy at the SUSY 
breaking minimum is set to zero as in Eq.~(\ref{eq:btypesbmin}). 
Such parameter region of $\mu_B$ and $\epsilon$ allowing the 
SUSY solution is shown in Fig.~\ref{fig:1}. 
In the shaded region, the SUSY solution 
(\ref{eq:newsolgen}) is not allowed and the SUSY 
breaking solution does not become metastable due to the 
R-breaking effect represented by $\epsilon$. 
Note that we find no allowed region along the $\epsilon = 0$ 
axis, which corresponds to the case of the solution 
(\ref{eq:newsol}).

\subsection{Generic case: $g_1(\phi_i) \ne 0$}
\label{ssec:typebgen}
Finally we introduce nonvanishing $g_1(\phi)$. As in Sec.~5.2, 
we first consider the case with $m_1^i=0$ in Eq.~(\ref{eq:hx1phi2}). 
In this case we can still integrate $\phi_i$ and $X_{a \ne 1}$ by 
use of $D_{\phi_i}W=D_{X_{a \ne 1}}W=0$ 
resulting in $\phi_i=X_{a \ne 1}=0$. 

The remnant of these heavy fields would be the one-loop mass 
$m_X$ for $X_1=X$ in Eq.~(\ref{eq:qcorsugra}). The effective 
scalar potential is in the same form as Eq.~(\ref{eq:qcorsugra}) 
but the effective superpotential $W$ in $G=K+ \ln |W|^2$ is now 
replaced by Eq.~(\ref{eq:qdcpl}). 
For $\epsilon \ll c \sim f \sim m_X \ll 1$ in the unit with $M_{Pl}=1$, 
we can obtain a SUSY breaking Minkowski minimum 
\begin{eqnarray}
X_{SB} &=& \frac{2fc}{m_X^2}\,\big( 1+{\cal O}(\epsilon^2) \big), 
\label{eq:btypesbminmx}
\end{eqnarray}
where the R-breaking constant 
$$
c \ = \ f/\sqrt{3}+{\cal O}(f^3/m_X^2;\,\epsilon^2), 
$$
is determined by the vanishing vacuum energy condition. 

The SUSY ground state in the hyperplane 
$\phi_i=X_{a \ne 1}=0$ which originates from the R-breaking 
parameter $\epsilon$ is the same as Eq.~(\ref{eq:btypessmin}), 
and the above breaking minimum becomes metastable. 
Unlike (\ref{eq:btypesbmin}), the SUSY breaking minimum 
(\ref{eq:btypesbminmx}) is not affected by the R-breaking term 
at ${\cal O}(\epsilon)$ due to the one-loop mass $m_X$, that is, 
the SUSY minimum (\ref{eq:btypessmin}) is independent 
of the SUSY breaking minimum (\ref{eq:btypesbminmx}) 
at this order. There might exist SUSY points 
analogous to Eq.~(\ref{eq:newsolgen}) outside the hypersurface 
$\phi_i=X_{a \ne 1}=0$ also in this case, but the solution 
would be more complicated due to the nonvanishing $h_1^{ij}$ 
in Eq.~(\ref{eq:hx1phi2}). 

Finally we comment about the case with $m_1^i \ne 0$ in 
Eq.~(\ref{eq:hx1phi2}). In this case, as mentioned in 
Sec.~\ref{sec:atypegen}, the field $X_1$ has a SUSY 
mass with the same magnitude as those of $\phi_i$'s as shown in 
Eq.~(\ref{eq:g1mass}). Then the field $X_1$ in the field basis 
(\ref{eq:orgsprd}) is no longer special. In this generalized OR model 
with most general R-breaking terms, the total superpotential 
would be written as 
\begin{eqnarray}
W &=& fX_1 +\sum_{a=1} g_a(\phi_i)X_a 
+c(\phi_i)
+\frac{1}{2} \sum_{a,b=1} m 
\epsilon_{ab}(\phi_i) X_a X_b +\cdots 
\nonumber \\ &=& 
c+fX_1+\frac{1}{2}\mu_{IJ}\Phi_I \Phi_J +\cdots, 
\nonumber
\end{eqnarray}
where $\Phi_I=(X_a,\phi_i)$, $I=(a,i)$ and the ellipses 
denote the higher order terms in $\Phi_I$. 
The SUSY mass matrix $\mu_{IJ}$ is given by 
the R-breaking components, 
$\mu_{ab}=m\epsilon_{ab}(0)$, 
$\mu_{ij}= \partial_{\phi_i} \partial_{\phi_j} c(0)$ and 
the R-symmetric components, 
$\mu_{ai}=2 \partial_{\phi_i} g_a(0)$. 
Note that $\mu_{1i}=2\partial_{\phi_i} g_1(0)=2m_1^i$. 
After the unitary rotation which makes $\mu_{IJ}$ diagonal, 
the above superpotential takes the form of 
$$
W \ = \ 
c+fU_{1I}\Phi_I+\frac{1}{2}\mu_{I} \Phi_I^2 +\cdots, 
$$
where $U_{IJ}$ is the rotation matrix and 
$\mu_I$ represents the eigenvalues of $\mu_{IJ}$. 
The F-flat conditions, $D_IW=W_I+K_IW=0$, allow a solution 
in general and SUSY would not be broken for 
$m_1^i \sim f$.

\section{Conclusion}
\label{sec:conclusion}
We considered $N=1$ global and local supersymmetric models 
with a continuous global $U(1)_R$ symmetry, and studied 
the effect of explicit R-symmetry breaking terms in detail. 

In global supersymmetric models, based on the argument by ISS, 
we have shown that some sort of explicit R-symmetry breaking 
terms can restore SUSY, and the original SUSY breaking vacuum 
can become metastable when a certain (but not generic) class 
of explicit R-symmetry breaking terms are added and/or loop 
effects stabilize the original SUSY breaking minimum. 

We have executed similar analyses in R-symmetric supergravity models. 
First we examined the general argument by NS in supergravity and 
found that it also holds with local SUSY except for the  
nontrivial case where the K\"ahler potential allows solution for 
the second condition in Eq.~(\ref{eq:susyvc2}). We presented 
concrete examples of this exception.
These models lead to AdS SUSY 
stationary solutions and associated SUSY breaking vacua with 
lower vacuum energy.
We found the general argument that this 
class of SUSY solutions corresponds to at best a saddle point, 
referring to Appendix~\ref{app:rmass}. 

Then, we studied the generalized OR model in supergravity with 
explicit R-symmetry breaking terms. We analyzed the structure of 
newly appeared SUSY stationary points as a consequence of the 
R-breaking effect and classified them. We have shown that these 
SUSY solutions disappear for type-A breaking terms (\ref{eq:arb}), 
when we tune the R-breaking constant term in the superpotential 
such that the original SUSY breaking minimum has 
a vanishing vacuum energy. In this sense, the introduction of 
explicit R-breaking terms do not always lead to a metastability 
of the SUSY breaking vacuum. 
On the other hand, the introduction of type-B breaking terms 
(\ref{eq:brb}) could cause a metastability of SUSY Minkowski 
minimum. We examined a parameter region which yields metastable 
vacuum in some concrete examples.

\subsection*{Acknowledgement}
The authors would like to thank K.~Choi and T.~Higaki 
for useful discussions.
H.~A.\/ and T.~K.\/ are supported in part by the
Grand-in-Aid for Scientific Research \#182496 
and \#17540251, respectively.
T.~K.\/ is also supported in part by 
the Grant-in-Aid for
the 21st Century COE ``The Center for Diversity and
Universality in Physics'' from the Ministry of Education, Culture,
Sports, Science and Technology of Japan.

\appendix
\section{Supersymmetric masses involving R-axion}
\label{app:rmass}
In this appendix, we show some general results for the 
SUSY masses for the scalar component of an 
R-axion multiplet. For this analysis, it is convenient 
to redefine the R-charged superfield $Y$ by 
\begin{eqnarray}
R &=& \frac{2}{q_Y} \ln Y, 
\nonumber
\end{eqnarray}
where $R$ can be interpreted as the R-axion superfield. 
(Note that $R=-aT$ in Eq.~(\ref{eq:def:t}).) 
In this basis, the K\"ahler potential and the superpotential 
(\ref{eq:rsp1}) is written as 
\begin{eqnarray}
K &=& K(R+\bar{R},\chi_i,\bar\chi_i), 
\nonumber \\
W &=& e^R \zeta(\chi_i). 
\label{eq:rsp4}
\end{eqnarray}
{}From Eq.~(\ref{eq:rsp4}), we find 
$W^{-1}\partial_Y^m W=1$ where $m=1,2,.\ldots$, and obtain 
\begin{eqnarray}
G_{RR} &=& K_{RR}+W^{-1}W_{RR}-(W^{-1}W_R)^2 
\ = \ K_{RR} \ = \ K_{R\bar{R}}, 
\nonumber \\
G_{\chi_i R} &=& K_{\chi_i R}+W^{-1}W_{\chi_i R}
-(W^{-1}W_{\chi_i})(W^{-1}W_R) 
\ = \ K_{\chi_i R} \ = \ K_{\chi_i \bar{R}}. 
\nonumber
\end{eqnarray}
Substituting these into the general formulae for the second 
derivatives at the SUSY point, 
\begin{eqnarray}
V_{I\bar{J}} \Big|_{D_KW=0} 
&=& e^G (G^{M \bar{N}}G_{MI}G_{\bar{N}\bar{J}}-2G_{I\bar{J}}), 
\nonumber \\
V_{IJ} \Big|_{D_KW=0} 
&=& -e^G G_{IJ}, 
\nonumber
\end{eqnarray}
we find 
\begin{eqnarray}
V_{R\bar{R}} \Big|_{D_KW=0} &=& 
V_{RR} \Big|_{D_KW=0} \ = \ -K_{R\bar{R}}\,m_{3/2}^2, 
\label{eq:vrr} \\
V_{\chi_i \bar{R}} \Big|_{D_KW=0} &=& 
V_{\chi_i R} \Big|_{D_KW=0} \ = \ -K_{\chi_i \bar{R}}\,m_{3/2}^2, 
\label{eq:vpr}
\end{eqnarray}
where $m_{3/2}^2=e^G$ is the gravitino mass square.

{}From Eq.~(\ref{eq:vrr}), the mass squared eigenvalues of 
$({\rm Re}\,R,\,{\rm Im}\,R)$ can be calculated as 
$0$ and $-2m_{3/2}^2$ with the canonical kinetic terms 
normalized by $K_{R\bar{R}}>0$. 
The first massless eigenmode corresponds to the R-axion scalar 
associated to the spontaneously broken global $U(1)_R$ symmetry. 
The second negative-definite eigenvalue indicates that the special 
SUSY solution (\ref{eq:susyvc2}) is at best a saddle point 
solution. Note that the gravitino mass $m_{3/2}$ is nonvanishing at 
this point and the vacuum energy is negative. We also find from 
Eq.~(\ref{eq:vpr}) that the mixing-mass between $R$ and $\chi_i$ 
is vanishing if the K\"ahler (kinetic) mixing is vanishing, 
$K_{\chi_i \bar{R}}=0$. In this case, the R-axion direction is 
completely separated from the other fields $\chi_i$, that is,
the above mass eigenvalues of R-axion multiplet become exact 
in this case. 

Finally we comment that the second derivatives (\ref{eq:vrr}) and 
(\ref{eq:vpr}) are all vanishing at the SUSY point (\ref{eq:susyvc1}) 
where $m_{3/2}=0$. In this case, both the real and the imaginary 
scalar component of R-axion multiplet remain massless. 
Note that Eq.~(\ref{eq:susyvc2}) may also allow a solution 
even in this case if $\zeta$ is {\it not} a generic function.

\end{document}